\documentclass[12pt]{article}
\usepackage{times}

\usepackage{epsf}
\def\beq{\begin{equation}}
\def\eeq{\end{equation}}
\def\bea{\begin{eqnarray}}
\def\eea{\end{eqnarray}}

\def\nnb{\nonumber}

\def\rar{\rightarrow}
\def\nnb{\nonumber}

\def\ba{\begin{array}}
\def\ea{\end{array}}
\def\bea{\begin{eqnarray}}
\def\eea{\end{eqnarray}}
\def\BXll{$B \rightarrow \,X_d\, \ell^+ \ell^-$}

\title{Standard Model CP violation in \BXll  decays}
\author{\vspace{1cm}\\
         {\bf Zeynep Deniz Eygi}
         \, \, and \, \,
        {\bf G\"{u}rsevil  Turan}
        \thanks{E-mail address: gsevgur@metu.edu.tr}\,\,}
     \date{}
\begin{document}
\setlength{\baselineskip}{24pt} \maketitle
\setlength{\baselineskip}{7mm}
~~~~~~~~~~~~~~~~~~~~~~~~~~{\it Middle East Technical University, Physics Dept. Inonu Bul.

~~~~~~~~~~~~~~~~~~~~~~~~~~~~~~~~~~~~~~~~~~~~~~~~~~ 06531 Ankara, TURKEY}

\abstract{We investigate the CP violating asymmetry, the forward backward asymmetry and the CP
violating asymmetry in the forward-backward asymmetry for the inclusive \BXll decays for
the $\ell=~e,\,\mu,\,\tau$ channels in the standard model. It is
observed that these asymmetries are quite sizable and   \BXll decays seem promising for
investigating CP violation.}
\thispagestyle{empty} \setcounter{page}{1}
\section{Introduction}
An efficient way in performing the precision test  for the
standard model (SM) is provided by the flavor-changing neutral
current (FCNC) processes since these are generated only through
higher order loop effects   in weak interaction. Among them, the
inclusive $B \rar X_{s(d)} \ell^+ \ell^-$ modes are  prominent
because of their relative cleanness compared to the pure hadronic
decays. In the SM, $B \rar X_{s(d)} \ell^+ \ell^-$ decays are
dominated by the parton level processes $b \rar s(d) \ell^+ \ell^-$,
which occur through an intermediate $u$, $c$ or $t$ quarks. They
can be described in term of an effective Hamiltonian which
contains the information about the short and long distance
effects.

The FCNC decays are also relevant to the CKM phenomenology; and $b \rar d \ell^+ \ell^-$
modes are especially important in this respect.
In case of the $b \rar s \ell^+ \ell^-$ decays, the
matrix element receives a combination of various contributions
from the intermediate  $t$, $c$ or $u$ quarks with factors
$V_{tb}V^*_{ts}\sim \lambda^2 $, $V_{cb}V^*_{cs}\sim \lambda^2$
and $V_{ub}V^*_{us}\sim \lambda^4$, respectively, where $\lambda =\sin \theta_C
\cong 0.22$. Since the last factor is extremely
small compared to the other two we can neglect it and this  reduces  the unitarity relation
for the CKM factors  to the form
$V_{tb}V^*_{ts}+V_{cb}V^*_{cs}\approx 0$. Hence, the matrix
element for the $b \rightarrow s\ell^+\ell^-$ decays involve only
one independent CKM factor so that  CP violation would not
show up. On the other hand, as pointed out before \cite{Kruger,Choudhury1}, for $b
\rightarrow d \ell^+\ell^-$ decay, all the CKM factors
$V_{tb}V^*_{td}$, $V_{cb}V^*_{cd}$ and $V_{ub}V^*_{ud}$ are at the
same order $\lambda^3$ in the SM  and the matrix element for these
processes would have sizable interference terms, so as to  induce a
CP violating asymmetry between the decay rates of the reactions $b
\rightarrow d \ell^+\ell^-$ and $\bar{b}\rightarrow
\bar{d}\ell^+\ell^-$. Therefore, $b \rightarrow d \ell^+\ell^-$ decays
seem to be suitable for establishing CP violation in B mesons.

 We note that the inclusive $B \rar X_{s} \ell^+ \ell^-$
decays have been widely studied in the framework of the SM and its
various extensions \cite{Hou}-\cite{Ergur}. As for $B \rar X_{d} \ell^+ \ell^-$ modes,
they were first considered within the SM  in \cite{Kruger}
and \cite{Choudhury1}. In ref. \cite{Kruger}, together with the branching ratio,
the CP violating asymmetry for the \BXll decays has been studied including the long-distance (LD)
 effects, but only for $\ell =e$ mode. In \cite{Choudhury1}, a SM  analysis for the forward-backward
asymmetry is given again only for $\ell = e$ mode and neglecting the LD contributions.
The general
two Higgs doublet model contributions and minimal supersymmetric extension of
the SM (MSSM) to the CP asymmetries were discussed in refs. \cite{Aliev1}
and \cite{Choudhury2}, respectively. Ref.  \cite{Choudhury2} contains a comparative study of the CP
asymmetries in the inclusive \BXll and exclusive $B \rightarrow \,\gamma \, \ell^+ \ell^-$ decays
for $\ell = \tau $ only, by mainly focusing on the effects of the scalar interactions in
the framework of the MSSM. Recently, CP violation
in the polarized $b \rar d \ell^+ \ell^-$ decay has been also
investigated in the SM \cite{Babu} and also in a general model
independent way \cite{Aliev2}. The aim of this work is to perform a quantitative analysis on the
SM CP violation and the related observables, such as the
forward-backward asymmetry and CP violation aysmmetry in the
forward-backward asymmetry in the $B \rar X_{d} \ell^+ \ell^-$
decays, some of which have already addressed in refs.
\cite{Kruger}, \cite{Choudhury1} and \cite{Choudhury2}, as pointed out above. However,
in this work we extend the investigation of the abovemensioned observables to consider all three
lepton modes  by  mainly focusing on LD effects and also their dependence on the  SM parameters
$\rho$ and $\eta$.

From the experimental side,  the
branching ratio $(BR)$ of the $B \rar X_{s} \ell^+ \ell^-$ decay
has been also reported by the BELLE Collaboration \cite{Belle},
$BR(B \rar X_{s} \ell^+ \ell^-)=((6.1 \pm 1.4)^{+1.4}_{-1.1})$, which is very close to the value
predicted by the SM \cite{Ali}, and may be used to put further constraint
on  the models beyond the SM.

We organized the paper as follows: Following this  brief introduction,
in section \ref{sect1}, we  first present the effective
Hamiltonian. Then, we introduce  the basic formulas
of the double and differential  decay rates,  CP violation asymmetry, $A_{CP}$,
forward-backward asymmetry, $A_{FB}$, and CP violating asymmetry in
forward-backward asymmetry $A_{CP}(A_{FB})$ for \BXll decay.
Section \ref{sect2} is devoted to the numerical analysis and discussion.

\section{The theoretical framework of $B \rar X_{d} \ell^+ \ell^-$ decays}\label{sect1}
Inclusive decay rates of the heavy hadrons can be calculated in the heavy quark effective theory
(HQET) \cite{HQET} and the important result from this procedure is that the leading terms in
$1/m_q$ expansion turn out to be the decay of a free quark, which can be calculated in the
perturbative QCD; while the corrections to the partonic decay rate start with $1/m^2_q$ only.
On the other hand, the powerful framework for both the inclusive and the exclusive modes
into which the perturbative QCD corrections to the physical decay amplitude  are incorporated
in a systematic way is the effective Hamiltonian method. In this approach, heavy degrees of freedom,
namely $t$ quark and $W^{\pm}$ bosons in the present case, are integrated out. The procedure
is to take into account the QCD corrections through matching the full theory with the
effective low energy one at the high scale $\mu =m_W$ and evaluating the Wilson coefficients from $m_W$
down to the lower scale $\mu \sim {\cal O}(m_b)$.
The effective
Hamiltonian obtained in this way for the
process $b \rar d \,  \ell^+ \ell^-$, is given by
\cite{Buras}, \cite{Buchalla}-\cite{Misiak}:
\begin{eqnarray}\label{Hamiltonian} {\cal H}_{eff}  =  \frac{4
G_F}{\sqrt{2}} \, V_{tb} V^*_{td}\Bigg\{ \sum_{i=1}^{10}& \,
C_i (\mu ) \, O_i(\mu)-\lambda_u
\{C_1(\mu)[O_1^u(\mu)-O_1(\mu)]\nnb\\&+C_2(\mu)[O_2^u(\mu)-O_2(\mu)]\}\Bigg\}
\end{eqnarray}
where
\bea\label{CKM}
\lambda_u=\frac{V_{ub}V_{ud}^\ast}{V_{tb}V_{td}^\ast},\label{lamu}
\eea
using the unitarity of the CKM matrix i.e.
$V_{tb}V_{td}^\ast+V_{ub}V_{ud}^\ast=-V_{cb}V_{cd}^\ast$. The
explicit forms of the operators $O_i$ can be found in
refs. \cite{Buchalla,Wise}. In Eq.(\ref{Hamiltonian}),
$C_i(\mu)$ are the Wilson coefficients calculated at a
renormalization point $\mu$ and their evolution from the higher scale $\mu=m_W$
down to the low-energy scale $\mu=m_b$ is described by the renormalization group
equation. For $C^{eff}_7(\mu)$ this calculation is performed in refs.\cite{Borzumati,Ciu}
in next to leading order. The value of $C_{10}(m_b)$ to the leading logarithmic approximation
can be found e.g. in
\cite{Buchalla,Misiak}. We here present the expression for
$C_9(\mu)$ which contains the terms responsible for the CP
violation in \BXll decay. It has a perturbative
part and a part coming from long distance (LD) effects due to conversion of the
real $\bar{c}c$ into lepton pair $\ell^+ \ell^-$:
\begin{eqnarray}
C_9^{eff}(\mu)=C_9^{pert}(\mu)+ Y_{reson}(s)\,\, ,
\label{C9efftot}
\end{eqnarray}
where
\begin{eqnarray}\label{Cpert}
C_9^{pert}(\mu)&=& C_{9}+h(u,s) [ 3 C_1(\mu) + C_2(\mu) + 3
C_3(\mu) + C_4(\mu) + 3 C_5(\mu) + C_6(\mu) \nonumber
\\&+&\lambda_u(3C_1 + C_2) ] -  \frac{1}{2} h(1, s) \left( 4
C_3(\mu) + 4 C_4(\mu)
+ 3 C_5(\mu) + C_6(\mu) \right)\nnb \\
&- &  \frac{1}{2} h(0,  s) \left[ C_3(\mu) + 3 C_4(\mu) +\lambda_u
(6 C_1(\mu)+2C_2(\mu)) \right] \\&+& \frac{2}{9} \left( 3 C_3(\mu)
+ C_4(\mu) + 3 C_5(\mu) + C_6(\mu) \right) \nonumber \,\, ,
\end{eqnarray}
and
\begin{eqnarray}
Y_{reson}(s)&=&-\frac{3}{\alpha^2}\kappa \sum_{V_i=\psi_i}
\frac{\pi \Gamma(V_i\rightarrow \ell^+
\ell^-)m_{V_i}}{m_B^2 s-m_{V_i}+i m_{V_i}
\Gamma_{V_i}} \nonumber \\
&\times & [ (3 C_1(\mu) + C_2(\mu) + 3 C_3(\mu) + C_4(\mu) + 3
C_5(\mu) + C_6(\mu))\nnb\\ &+&\lambda_u(3C_1(\mu)+C_2(\mu))]\, .
 \label{Yresx}
\end{eqnarray}
In Eq.(\ref{Cpert}), $s=q^2/m_B^2$ where $q$ is the momentum transfer, $u=\frac{m_c}{m_b}$
 and the functions $h(u, s)$ arise from one loop
contributions of the four-quark operators $O_1-O_6$ and are given by
\begin{eqnarray}
h(u, s) &=& -\frac{8}{9}\ln\frac{m_b}{\mu} - \frac{8}{9}\ln u +
\frac{8}{27} + \frac{4}{9} y \\
& & - \frac{2}{9} (2+y) |1-y|^{1/2} \left\{\begin{array}{ll}
\left( \ln\left| \frac{\sqrt{1-y} + 1}{\sqrt{1-y} - 1}\right| -
i\pi \right), &\mbox{for } y \equiv \frac{4u^2}{ s} < 1 \nonumber \\
2 \arctan \frac{1}{\sqrt{y-1}}, & \mbox{for } y \equiv \frac
{4u^2}{ s} > 1,
\end{array}
\right. \\
h(0,s) &=& \frac{8}{27} -\frac{8}{9} \ln\frac{m_b}{\mu} -
\frac{4}{9} \ln s + \frac{4}{9} i\pi \,\, . \label{hfunc}
\end{eqnarray}
The phenomenological parameter $\kappa$
in Eq. (\ref{Yresx}) is taken as $2.3$ (see e.g. \cite{Ligeti}).

The next step is to calculate the matrix element of the $B \rar X_{d} \ell^+ \ell^-$ decay.
Neglecting the mass of the $d$ quark,  the effective short distance Hamiltonian
 in Eq.(\ref{Hamiltonian})  leads to the following  QCD corrected matrix element:
\begin{eqnarray}\label{genmatrix}
{\cal M} &=&\frac{G_{F}\alpha}{2\sqrt{2}\pi }V_{tb}V_{td}^{\ast }%
\Bigg\{C_{9}^{eff}(m_{b})~\bar{d}\gamma _{\mu }(1-\gamma _{5})b\,\bar{\ell}%
\gamma ^{\mu }\ell +C_{10}(m_{b})~\bar{d}\gamma _{\mu }(1-\gamma _{5})b\,\bar{%
\ell}\gamma ^{\mu }\gamma _{5}\ell  \nonumber \\
&-&2C_{7}^{eff}(m_{b})~\frac{m_{b}}{q^{2}}\bar{d}i\sigma _{\mu \nu
}q^{\nu }(1+\gamma _{5})b\,\bar{\ell}\gamma ^{\mu }\ell
\Bigg\}.
\end{eqnarray}
Since the initial and final state polarizations are not measured, we must average over
the initial spins and sum over the final ones, that leads to the following double differential
decay rate
\bea
\frac{d^2 \Gamma}{d s \, dz} & = & \Gamma(B \to X_c \ell \nu) \frac{\alpha^2 }{4 \pi^2 f(u) k(u)}  (1 - s)^2
    \frac{|V_{tb} V^{\ast}_{td}|^2}{|V_{cb}|^2} \, v \,\Big \{ 12\, v \, z \,
\mbox{\rm Re}(C^{eff}_{7} C^{*}_{10}) \nnb \\
& + & 12 \, \Big(1 + \frac{2 t}{s}\Big ) \,  \mbox{\rm Re}(C^{eff}_{7} C^{eff \,*}_{9})+
6 \, v \, \mbox{\rm Re}(C_{10} C^{eff \,*}_{9}) \nnb \\
& + & \frac{3}{2} \Big [(1+s)-(1-s)\, v^2 z^2+4 t \Big ]|C^{eff }_{9}|^2 \nnb \\
& + & 6 \, \Big [\Big (1+\frac{1}{s}\Big )-\Big (1-\frac{1}{s}\Big )\, v^2 z^2+\frac{4 t}{s}
 \Big ]|C^{eff }_{7}|^2 \nnb \\
& + & \frac{3}{2} \Big [(1+s)-(1-s) \,v^2 z^2-4 t \Big ]|C_{10}|^2 \Big \} \label{DD}
\eea
where $ v=\sqrt{1 - 4 t/s}$, $t=m^2_{\ell}/m^2_b$  and $z=\cos \theta$, where
$\theta$ is the angle between the momentum of the B-meson and that of $\ell^-$
in the center of mass frame of the dileptons $\ell^-\ell^+$. In Eq. (\ref{DD}),
\beq
\Gamma(B \to X_c \ell \nu) = \frac{G_F^2 m_b^5}{192 \pi^3} |V_{cb}|^2 f(u) k(u) \, ,
\eeq
where
\bea
f(u) &=& 1 - 8 u + 8 u^4 - u^8 - 24 u^4 ln (u)   \\
k(u) &=& 1 - \frac{2 \alpha_s(m_b)}{3 \pi}
    \Bigg[ \left( \pi^2 - \frac{31}{4} \right)(1 - \hat{m}_c^2) + {3 \over 2} \Bigg] \, ,
\eea
are the phase space factor and the QCD corrections to the semi-leptonic decay rate, respectively,
which is used to normalize the decay rate of $B \rar X_{d} \ell^+ \ell^-$
to remove the uncertainties in the value of $m_b$.

After integrating the double differential decay rate in Eq.(\ref{DD}) over the angle
variable, we find
\beq
\frac{d \Gamma}{d s}
  = \Gamma(B \to X_c \ell \nu) \frac{\alpha^2 }{4 \pi^2 f(u) k(u)}  (1 - s)^2
    \frac{|V_{tb} V^{\ast}_{td}|^2}{|V_{cb}|^2} \sqrt{1 - \frac{4 t}{s}} \, \Delta (s)\, ,
\label{rate}
\eeq
where
\bea
\Delta (s) & = & \frac{(s+2 s^2+2 t-8 s t)}{s}|C_{10}|^2+\frac{4}{s^2}(2+s)(s+2 t)|C^{eff}_{7}|^2
+(2+s)(1+ \frac{2 t}{s})|C^{eff}_{9}|^2 \nnb \\
&+ & \frac{12}{s}(s+2 t) \mbox{\rm Re}(C^{eff}_{7} C^{eff \,*}_{9})\, .
\eea

We start with calculating the CP  asymmetry $A_{CP}$ between  the $B \rar X_{d} \ell^+ \ell^-$
and the conjugated one $\bar{B} \rar \bar{X_{d}} \ell^+ \ell^-$, which is defined as
\beq
A_{CP}(s) ~=~ \frac{\frac{d \Gamma}{ds} - \frac{d
\bar{\Gamma}}{d s}}{\frac{d \Gamma}{d s} + \frac{d
\bar{\Gamma}}{d s}}
\label{ACPdef}
\eeq
where
\beq
\frac{d\Gamma}{d s} ~=~ \frac{d\Gamma(B \rar X_{d} \ell^+ \ell^-)}{d s} ~~,~~
\frac{d\bar{\Gamma}}{d s} ~=~ \frac{d\Gamma(\bar{B} \rar \bar{X_{d}} \ell^+ \ell^-)}{d s} \,.
\eeq
 Since in the SM only $C^{eff }_{9}$ contains imaginary part, representing $C^{eff}_9$ symbolically as
\beq
C^{eff}_9 ~=~ \xi_1 + \lambda_u  \xi_2
\label{C9}
\eeq
and further substituting $\lambda \rightarrow \lambda^{\ast}$ for the conjugated process
$\bar{B} \rar \bar{X_{d}} \ell^+ \ell^-$, one can easily obtain \cite{Kruger}
\bea
A_{CP} (s) & = & \frac{-2 \, \mbox{\rm Im}(\lambda_u)\,\Sigma }{\Delta +2 \, \mbox{\rm Im}(\lambda_u)\,\Sigma}
\, ,
\eea
where
\bea
\Sigma & = & \Big ( 1+ \frac{2 t}{s}\Big ) [ (1+2 s) \,\mbox{\rm Im}(\xi^{\ast}_1 \xi_2)+
6 \, C^{eff }_{7}\, \mbox{\rm Im}(\xi_2)] \, \mbox{\rm Im}(\lambda_u)\, .
\eea

For completeness, we next consider the forward-backward asymmetry, $A_{FB}$, in \BXll, which is another
physical quantity that may be useful to test the theoretical models. Using the definition of
differential $A_{FB}(s)$
\begin{eqnarray}
A_{FB}(s)& = & \frac{ \int^{1}_{0}dz \frac{d^2 \Gamma }{ds dz} -
\int^{0}_{-1}dz \frac{d^2 \Gamma }{ds dz}}{\int^{1}_{0}dz
\frac{d^2 \Gamma }{ds dz}+ \int^{0}_{-1}dz \frac{d^2 \Gamma }{ds dz}}\, ,
\label{AFB1}
\end{eqnarray}
we find
\begin{eqnarray}
A_{FB}(s) =\frac{3~ v }{\Delta (s)} \,\mbox{\rm Re}[C_{10}(2 C^{eff}_{7}+s \, C^{eff \,*}_{9})],
\label{AFB2}
\end{eqnarray}
which agrees with the result given by ref. \cite{Choudhury1}, but not by  \cite{Choudhury2}.

We have also a CP violating asymmetry in $A_{FB}$, $A_{CP}
(A_{FB})$, in \BXll decay.  Since in the limit of CP conservation, one expects $A_{FB}=-\bar{A}_{FB}$
\cite{Choudhury1,Buchalla2}, where
$A_{FB}$ and $\bar{A}_{FB}$ are the  forward-backward asymmetries in the particle and
antiparticle channels, respectively, $A_{CP}(A_{FB})$ is defined as
\begin{eqnarray}
A_{CP}(A_{FB})& = & A_{FB} +\bar{A}_{FB} ~~. \label{ACPAFB1}
\end{eqnarray}
Here, $\bar{A}_{FB}$  can be obtained by the replacement,
\bea
C_9^{eff}(\lambda_u)\rightarrow \bar{C}_9^{eff}(\lambda_u \rightarrow \lambda_u^\ast).
\eea
Using Eqs.(\ref{AFB2}) we can find
\bea
A_{CP}
(A_{FB}) & = & \frac{6 ~ v~\mbox{\rm Im}(\lambda_u)}{\Delta (\Delta +4 \mbox{\rm Im}(\lambda_u)~\Sigma)}~C_{10}
\nnb \\ & \cdot &
\Big [ 2 \Sigma ~(2 C^{eff}_7+s(\mbox{\rm Re}(\xi_1)+\mbox{\rm Re}(\xi_2)~\mbox{\rm Re}(\lambda_u)-
\mbox{\rm Im}(\xi_2)~\mbox{\rm Im}(\lambda_u)))-s ~ \Delta ~ \mbox{\rm Im}(\xi_2)  \Big ] \, , \nnb \\
\eea
which is slightly different from the one given by ref.  \cite{Choudhury2}.

\section{Numerical analysis and discussion \label{sect2}}
\begin{figure}[htb]
\vskip 0truein \centering \epsfxsize=3.8in
\leavevmode\epsffile{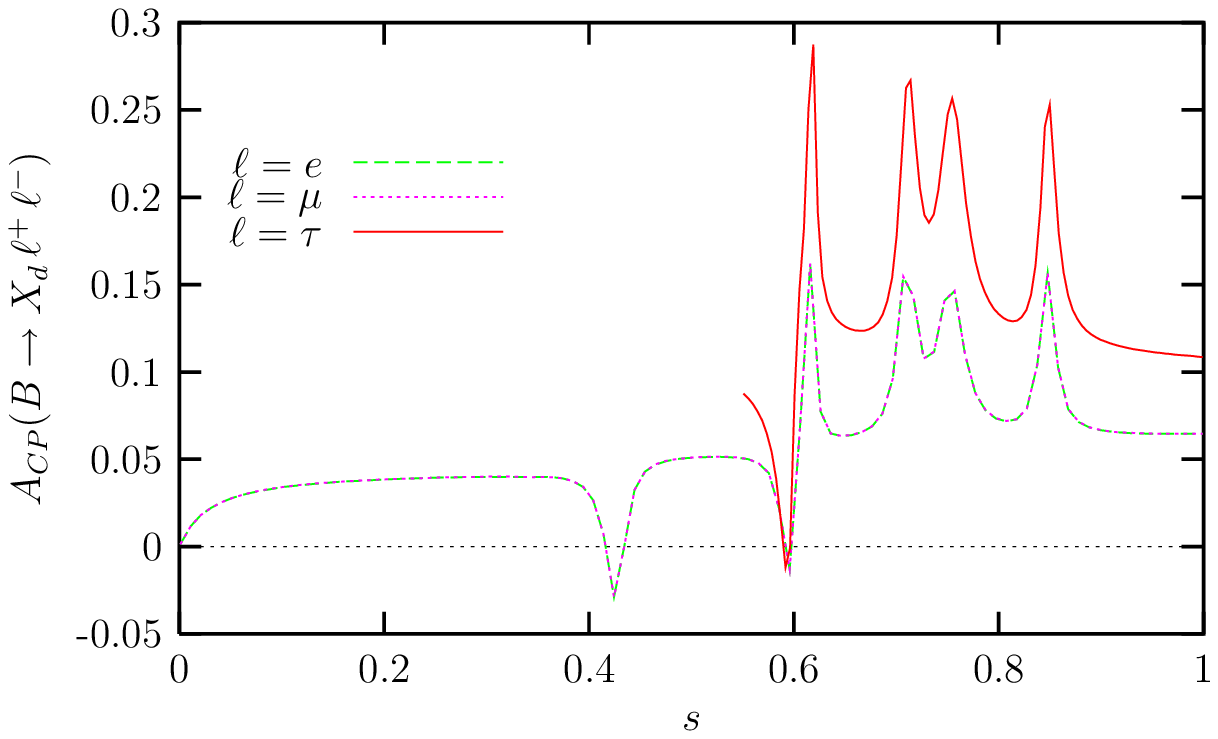} \vskip 0truein
\caption[]{$A_{CP}$ for \BXll decay for the Wolfenstein parameters
$(\rho,\,\eta)=(0.15;\,0.30)$. The three distinct lepton modes $\ell=~e,~\mu,~\tau$
are represented by the  dashed, dotted  and solid curves,
respectively. } \label{ACP3}
\end{figure}

\begin{figure}[htb]
\vskip 0truein \centering \epsfxsize=3.8in
\leavevmode\epsffile{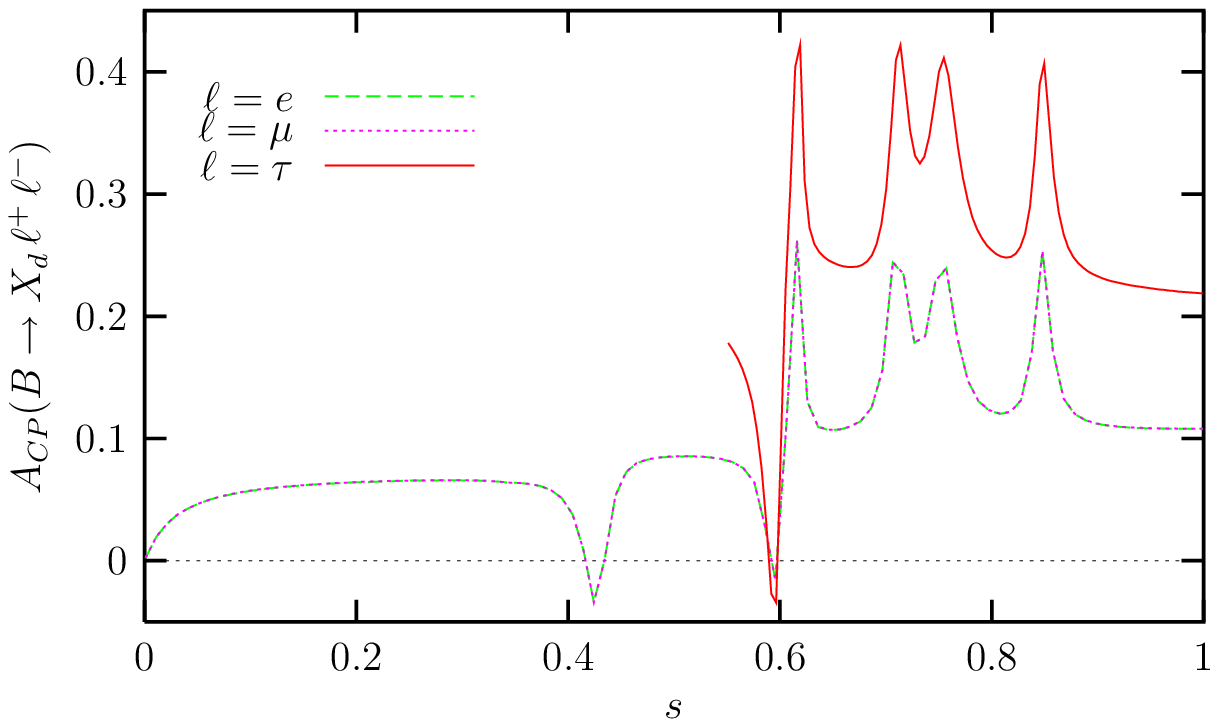} \vskip 0truein \caption[]{The same
as Fig.(\ref{ACP3}) but for the Wolfenstein parameters
$(\rho,\,\eta)=(0.32;\,0.38)$} \label{ACP5}
\end{figure}

\begin{figure}[htb]
\vskip 0truein \centering \epsfxsize=3.8in
\leavevmode\epsffile{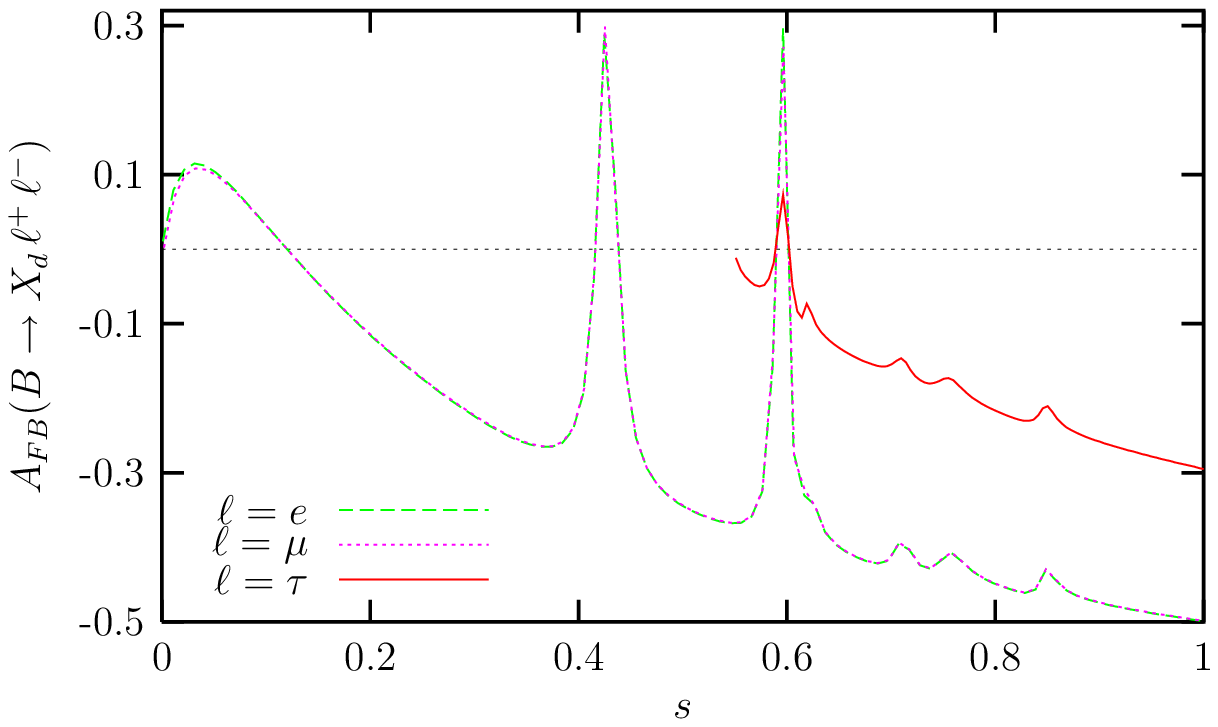} \vskip 0truein \caption[]{$A_{FB}$ for \BXll decay for the Wolfenstein
parameters $(\rho,\,\eta)=(0.15;\,0.30)$. The three distinct lepton modes $\ell=~e,~\mu,~\tau$
are represented by the  dashed, dotted and solid curves,
respectively.} \label{AFB3}
\end{figure}

\begin{figure}[htb]
\vskip 0truein \centering \epsfxsize=3.8in
\leavevmode\epsffile{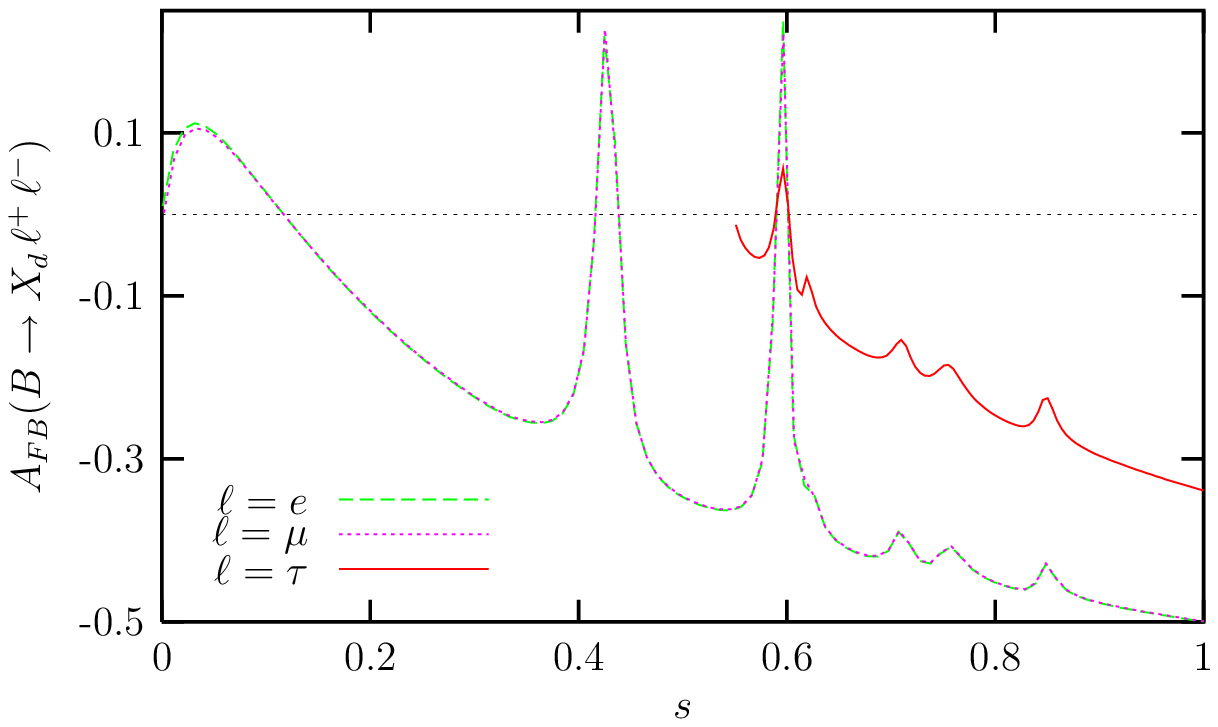} \vskip 0truein \caption[]{The same
as Fig.(\ref{AFB3}) but for the Wolfenstein parameters $(\rho,\,\eta)=(0.32;\,0.38)$} \label{AFB5}
\end{figure}

\begin{figure}[htb]
\vskip 0truein \centering \epsfxsize=3.8in
\leavevmode\epsffile{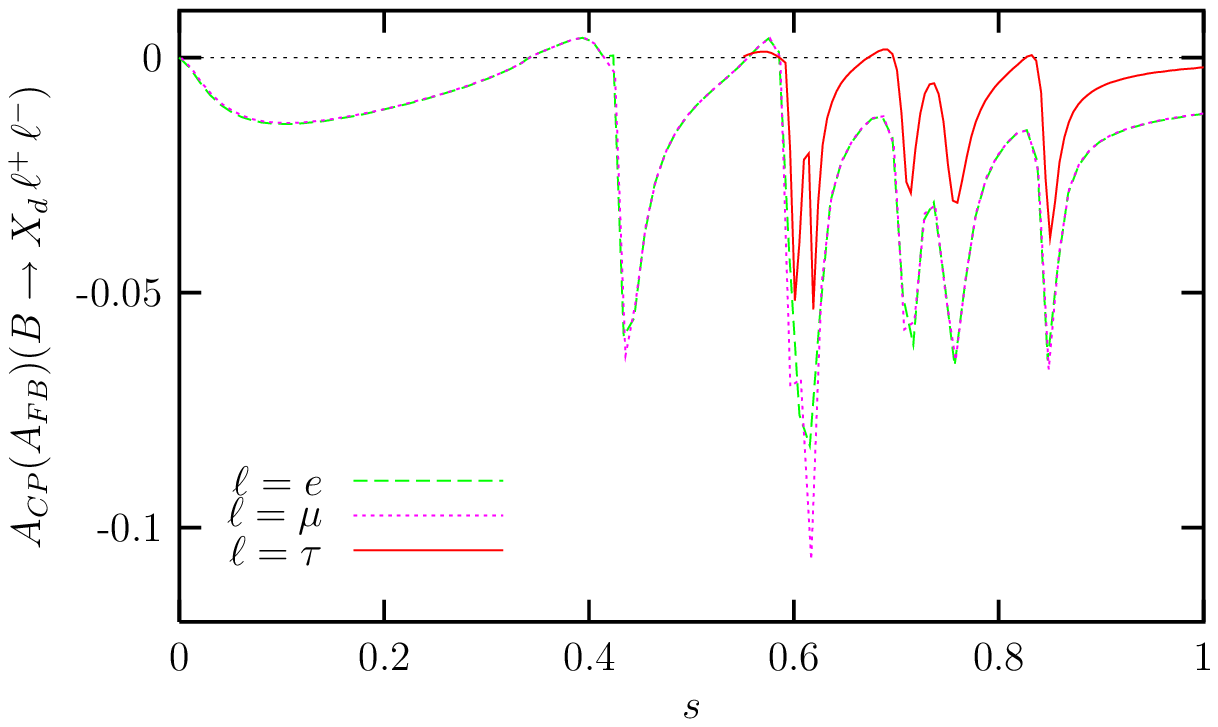} \vskip 0truein
\caption[]{$A_{CP}(A_{FB})$ for \BXll decay for the Wolfenstein
parameters $(\rho,\,\eta)=(0.15;\,0.30)$. The three distinct lepton modes $\ell=~e,~\mu,~\tau$
are represented by the dashed, dotted and solid curves, respectively.} \label{ACPAFB3}
\end{figure}

\begin{figure}[htb]
\vskip 0truein \centering \epsfxsize=3.8in
\leavevmode\epsffile{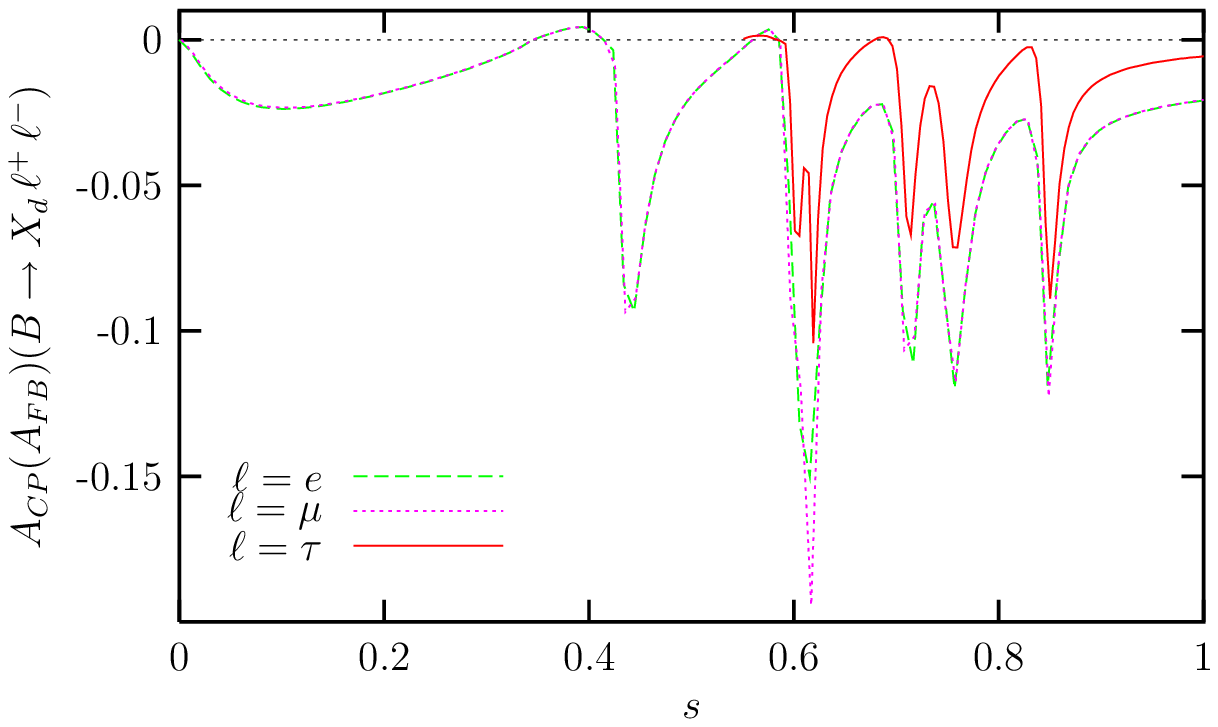} \vskip 0truein \caption[]{The
same as Fig.(\ref{ACPAFB3}) but for the Wolfenstein parameters
$(\rho,\,\eta)=(0.32;\,0.38)$} \label{ACPAFB5}
\end{figure}

In this section, we present results of our calculations related to \BXll decays, for two
different sets of the Wolfenstein parameters. For this we first give the
Wolfenstein parametrization \cite{Wolf} of the CKM factor in Eq.(\ref{lamu})
\bea
\lambda_u=\frac{\rho(1-\rho)-\eta^2-i\eta}{(1-\rho)^2+\eta^2}+O(\lambda^2)\, ,
\eea
and also
\bea
 \frac{|V_{tb} V^{\ast}_{td}|^2}{|V_{cb}|^2} & = & \lambda^2 [(1-\rho)^2+\eta^2]+
 {\cal O}(\lambda^4)\, .
\eea
The updated fitted values for the  parameters $\rho$ and $\eta$ are given in ref.\cite{AliLunghi} as
\bea
\bar{\rho} & = & 0.22 \pm 0.07 \, \, (0.25\pm 0.07)\nnb \\
\bar{\eta} & = & 0.34 \pm 0.04 \, \, (0.34\pm 0.04)\label{param}
\eea
with (without) including the chiral logarithms uncertainties. In our numerical analysis,
we have used  $(\rho,\,\eta)=(0.15;\,0.30)$ and $(0.32;\,0.38)$, which are the lower and higher allowed
values of the parameters given in Eq. (\ref{param}) above, and
 present the dependence of the $A_{CP}$, $A_{FB}$
and $A_{CP} (A_{FB})$ on the dimensionles photon energy $s$ for the \BXll $(\ell=e,\,\mu,\,\tau)$ decays
 in Figs. (\ref{ACP3}-\ref{ACPAFB5}).

We have also evaluated the average values of CP asymmetry $<A_{CP}>$, forward-backward
asymmetry $<A_{FB}>$ and CP asymmetry in the forward-backward asymmetry $<A_{CP} (A_{FB})>$
in \BXll decay for the above  sets of parameters $(\rho,\,\eta)$, and our results are displayed in Table 1 and
2 without and with including the long distance effects, respectively.

The input parameters and the initial values of the Wilson
coefficients we used in our numerical
analysis are as follows:
\begin{eqnarray}
& & m_B =5.28 \, GeV \, , \, m_b =4.8 \, GeV \, , \,m_c =1.4 \,
GeV \, ,m_t=175 \, GeV  \, , \nnb \\
& & m_e=0.511 \,MeV \, \, , \, \, m_{\tau} =1.777 \, GeV,\, m_{\mu}=0.105\,GeV,\, ,\,\nnb\\
& & BR(B\rightarrow X_c e \bar{\nu}_e)=10.4 \% \, \, , \, \, \alpha =1/129 \, \, , \, \,
m_W=80.4 \, GeV \, \, , \, \, m_Z=91.1 \, GeV \nnb \\
& &C_1=-0.245,\, C_2=1.107,\, C_3=0.011,\, C_4=-0.026,\,C_5=0.007,\,\nnb\\
& & C_{6}=-0.0314,\,C^{eff}_{7}=-0.315,\, C_{9}=4.220, \, C_{10}=-4.619.
\end{eqnarray}

In our numerical analysis, we take into account five possible
resonances for the LD effects coming from the reaction $b \rar d\,
\psi_i \rar d \, \ell^+ \ell^-$, where $i=1,...,5$ and divide the
integration region into two parts for $\ell=\tau$: $(2 m_{\ell}/m_B)^2 \leq s \leq
((m_{\psi_1}-0.02)/m_B)^2$ and $((m_{\psi_1}+0.02)/m_B)^2 \leq
s \leq 1 $, where $m_{\psi_1}=3.097$ GeV is
the mass of the first resonance. As for $\ell=e$ and $\mu$ modes, the
integration region is divided into three parts : $(2 m_{\ell}/m_B)^2 \leq s \leq
((m_{\psi_1}-0.02)/m_B)^2$, $((m_{\psi_1}+0.02)/m_B)^2 \leq s \leq ((m_{\psi_2}-0.02)/m_B)^2$ and
$((m_{\psi_2}+0.02)/m_B)^2 \leq s\leq 1 $, where $m_{\psi_2}=3.686 $ GeV is
the mass of the second resonance.

For reference, we present our SM predictions with long distance effects
\bea
BR( B\rightarrow X_d \, \ell^+ \, \ell^-) & = & (3.01,2.61,0.11)\times 10^{-7}\, ,
\eea
for $\ell=e,\mu,\tau$, respectively, with $(\rho;\eta)=(0.30;0.34)$, which is in agreement
with the results of ref.\cite{Kruger}.

\begin{table}
\begin{center}
\begin{tabular}{|c|ccc|ccc|ccc|}
 \hline\hline
  &&{\scriptsize$<A_{CP}>$}&&&{\scriptsize$<A_{FB}>$}&&&{\scriptsize$<A_{CP}(A_{FB}>$)}&\\
 {\scriptsize$(\rho;\,\eta)$}&{\scriptsize$\ell~=~e$}&{\scriptsize$\ell~=~\mu$}&{\scriptsize$\ell~=~\tau$}& {\scriptsize$\ell~=~e$}&{\scriptsize$\ell~=~\mu$}
 &{\scriptsize$\ell~=~\tau$}&{\scriptsize$\ell~=~e$}&{\scriptsize$\ell~=~\mu$}&{\scriptsize$\ell~=~\tau$}\\
\hline {\scriptsize$(0.15;\,0.30)$}&$0.030$&$0.036$&$0.134$&$-0.124$&$-0.151$&$-0.182$&$-0.009$&$-0.009$&$0.001$\\
\hline
{\scriptsize$(0.32;\,0.38)$}&$0.051$ & $0.061$&$0.169$&$-0.129$&$-0.156$&$-0.180$&$-0.015$&$-0.015$&$0.002$\\
  \hline\hline
  \end{tabular}
  \end{center}
  \caption{The average values of
$A_{CP}$, $A_{FB}$ and $A_{CP}(A_{FB})$ in \BXll for the three distinct
lepton modes without including the long distance effects.}\label{tab1}
  \end{table}
\begin{table}
\begin{center}
\begin{tabular}{|c|ccc|ccc|ccc|}
 \hline\hline
  &&{\scriptsize$<A_{CP}>$}&&&{\scriptsize$<A_{FB}>$}&&&{\scriptsize$<A_{CP}(A_{FB}>)$}&\\
 {\scriptsize$(\rho;\,\eta)$}&{\scriptsize$\ell~=~e$}&{\scriptsize$\ell~=~\mu$}&{\scriptsize$\ell~=~\tau$}&{ \scriptsize$\ell~=~e$}&{\scriptsize$\ell~=~\mu$}
 &{\scriptsize$\ell~=~\tau$}&{\scriptsize$\ell~=~e$}&{\scriptsize$\ell~=~\mu$}&{\scriptsize$\ell~=~\tau$}\\
\hline {\scriptsize$(0.15;\,0.30)$}&$0.032$&$0.036$&$0.144$&$-0.119$&$-0.139$&$-0.157$&$-0.017$&$-0.017$&$-0.004$\\
\hline
{\scriptsize $(0.32;\,0.38)$}&$0.051$ & $0.059$&$0.230$&$-0.125$&$-0.140$&$-0.150$&$-0.031$&$-0.030$&$-0.009$\\
  \hline\hline
  \end{tabular}
  \end{center}
  \caption{The same as Table (\ref{tab1}), but  including the long distance effects.}\label{tab2}
  \end{table}

In Fig.(\ref{ACP3}) and Fig.(\ref{ACP5}), we present the
dependence of $A_{CP}$ on the dimensionless photon energy $s$, for
\BXll decay for the Wolfenstein parameters
$(\rho;\,\eta)=(0.15;\,0.30)$ and $(\rho;\,\eta)=(0.32;\,0.38)$,
respectively. The three distinct lepton modes $\ell=~e,~\mu,~\tau$
are represented by the  dashed, dotted and solid curves,
respectively. We observe that the $A_{CP}$ for $\ell=~e,~\mu$
cases almost coincide, reaching  up to  $25~\%$  for the larger  values of $s$.
The $A_{CP}$ for $\ell=~\tau$ mode exceeds the values of the  other modes and
reaches $40~\%$. We also observe from Tables 1 and 2 that
 including the LD  effects in calculating $<A_{CP}>$ does not change the results
  for $\ell=~e,~\mu$ modes, while $\ell=~\tau$ mode, it is quite sizable,
 $8-36\%$, depending on the sets  of the parameters used for $(\rho;\,\eta)$.

The $s$ dependence of $A_{FB}$ for the \BXll $(\ell=~e,~\mu,~\tau)$ decays are plotted in
Figs.(\ref{AFB3}) and (\ref{AFB5}) for $(\rho;\,\eta)=(0.15;\,0.30)$ and $(\rho;\,\eta)=(0.32;\,0.38)$,
respectively.
We see that $A_{FB}$ is negative for almost all values of $s$, except in the resonance  and
very small-$s$ regions.
$<A_{FB}>$ takes the values between  $-(12-15)\%$ depending on the sets  of the parameters
used for $(\rho;\,\eta)$ . The LD effects on $<A_{FB}>$ are about $10\%$, but in reverse
manner, decreasing its magnitude in comparison to the values without LD contributions.

We present the dependence of the $A_{CP}(A_{FB})$ of \BXll decay
on  $s$ in Fig.(\ref{ACPAFB3}) and
Fig.(\ref{ACPAFB5}), again for two different sets of the
Wolfenstein parameters. As for $A_{CP}$,   $A_{CP}(A_{FB})$ for
$\ell=~e,$ and $\ell=~\mu$ modes almost coincide. We see that $A_{CP}(A_{FB})$ is all negative
except in a very small region for the intermediate  values of $s$ for $\ell=~e,~\mu$
cases. LD effects seem to be quite significant for $<A_{CP}(A_{FB})>$,
enhancing   its value twice (four times)  for $\ell=~e,~\mu$ ($\ell=~\tau$) modes.
To see this LD contributions more closely, we present the $<A_{CP}(A_{FB})>$
for different regions of  $s$ in Table (\ref{tab3}) and
(\ref{tab4}), for $(\rho;\,\eta)=(0.15;\,0.30)$ and $(\rho;\,\eta)=(0.32;\,0.38)$,
respectively. We see that for the light lepton modes, $\ell = e, \mu$,  $A_{CP}(A_{FB})$
is more sizable in the high-dilepton mass region of $s$, $((m_{\psi_2}+0.02)/m_B)^2 \leq s \leq 1$.
However, for $\ell =  \tau$, the contribution from the high-dilepton mass region of $s$ is
negligible and the contribution to $<A_{CP}(A_{FB})>$ comes effectively from the low-dilepton mass
region,  $(2m_l/m_B)^2\leq s \leq ((m_{\psi_1}-0.02)/m_B)^2$ and amounts to $-1 \%$.

\begin{table}
\begin{center}
\begin{tabular}{|c|c|c|c|c|c|}
 \hline\hline
   & {\scriptsize SD}   &
    $(2m_l/m_B)^2\leq s \leq $ & $((m_{\psi_1}+0.02)/m_B)^2\leq s $&
    $((m_{\psi_2}+0.02)/m_B)^2  $ & {\scriptsize SD+LD} \\
    $\ell$ & {\scriptsize contribution}& $((m_{\psi_1}-0.02)/m_B)^2$
    & $\leq ((m_{\psi_2}-0.02)/m_B)^2$ & $ \leq s \leq 1$ & {\scriptsize  contribution} \\
\hline\hline
e & $-0.92$ & $-0.29$ & $-0.25$ & $-1.20$ & $-1.78$ \\ \hline
$\mu$ & $-0.91$ & $-0.29$ & $-0.25$ & $-1.20$ & $-1.78$ \\ \hline
$\tau$ & $-0.11$ & $-0.42$ & $3.10\times 10^{-3}$ & & $-0.42$ \\ \hline\hline
  \end{tabular}
  \end{center}
  \caption{The SM predictions for the average CP-violating asymmetry in the forward-backward
  asymmetry $<A_{CP}(A_{FB})>\times 10^{-2}$
  for different regions of the dimensionless photon energy $s$ with $(\rho ; \eta )=(0.15;\,0.30)$.}
  \label{tab3}
\end{table}

\begin{table}
\begin{center}
\begin{tabular}{|c|c|c|c|c|c|}
 \hline\hline
   & {\scriptsize SD}   &
    $(2m_l/m_B)^2\leq s \leq $ & $((m_{\psi_1}+0.02)/m_B)^2\leq s $&
    $((m_{\psi_2}+0.02)/m_B)^2  $ & {\scriptsize SD+LD} \\
    $\ell$ & {\scriptsize contribution}& $((m_{\psi_1}-0.02)/m_B)^2$
    & $\leq ((m_{\psi_2}-0.02)/m_B)^2$ & $ \leq s \leq 1$ & {\scriptsize  contribution} \\
\hline\hline
e & $-1.59$ & $-0.51$ & $-0.43$ & $-2.15$ & $-3.10$ \\ \hline
$\mu$ & $-1.57$ & $-0.51$ & $-0.43$ & $-2.15$ & $-3.09$  \\ \hline
$\tau$ & $0.20$ & $-0.94$ & $3.30\times 10^{-3}$ & & $-0.94$ \\ \hline\hline
  \end{tabular}
  \end{center}
  \caption{Same as Table (\ref{tab3}), but with $(\rho ; \eta )=(0.32;\,0.38)$.}\label{tab4}
\end{table}
As a conclusion we can say that there is a significant $A_{CP}$
and $A_{CP}(A_{FB})$ for the \BXll decay, although the branching ratios
predicted for these channels are relatively small because of CKM
suppression. So,    \BXll decays seem promising for  investigating
CP violation.
\newpage

\newpage
\newpage
\newpage


\begin{thebibliography}{1}
\bibitem{Kruger} F. Kr\"{u}ger, L.M. Sehgal, {\it Phys. Rev.}, {\bf
D55}, (1997) 2799.
%
\bibitem{Choudhury1} S. Rai Choudhury, {\it Phys. Rev.}, {\bf D56}, (1997) 6028.
%
\bibitem{Hou} W. S. Hou, R. S. Willey and A. Soni,
{\it Phys. Rev. Lett.} {\bf 58} (1987) 1608.
%
\bibitem{R5} N. G. Deshpande and J. Trampetic,
{\it Phys. Rev. Lett.} {\bf 60} (1988) 2583.
%
\bibitem{R6} C. S. Lim, T. Morozumi and A. I. Sanda,
{\it Phys. Lett.} {\bf B218} (1989) 343.
%
\bibitem{Grinstein1} B. Grinstein, M. J. Savage and M. B. Wise,
{\it Nucl. Phys.} {\bf B319} (1989) 271.
%
\bibitem{R8} C. Dominguez, N. Paver and Riazuddin,
{\it Phys. Lett.} {\bf B214} (1988) 459.
%
\bibitem{R9} N. G. Deshpande, J. Trampetic and K. Ponose,
{\it Phys. Rev.} {\bf D39} (1989) 1461.
%
\bibitem{donnel} P. J. O'Donnell and H. K. Tung,
{\it Phys. Rev.} {\bf D43} (1991) 2067.
%
\bibitem{R12} N. Paver and Riazuddin,
{\it Phys. Rev.} {\bf D45} (1992) 978.

\bibitem{ali} A. Ali, T. Mannel and T. Morozumi,
{\it Phys. Lett.} {\bf B273} (1991) 505.

\bibitem{R14} A. Ali, G. F. Giudice and T. Mannel,
{\it Z. Phys.} {\bf C67} (1995) 417.

\bibitem{Greub} C. Greub, A. Ioannissian and D. Wyler,
{\it Phys. Lett.} {\bf B346} (1995) 145; \\
D. Liu {\it Phys. Lett.} {\bf B346} (1995) 355; \\
G. Burdman, {\it Phys. Rev.} {\bf D52} (1995) 6400: \\
Y. Okada, Y. Shimizu and M. Tanaka {\it Phys. Lett.} {\bf B405}
(1997) 297.
%
\bibitem{Buras} A. J. Buras and M. M\"{u}nz,
{\it Phys. Rev.} {\bf D52} (1995) 186.
%
\bibitem{R17} N. G. Deshpande, X. -G. He and J. Trampetic,
{\it Phys. Lett.} {\bf B367} (1996) 362.
%
\bibitem{Jaus} W. Jaus and D. Wyler,
{\it Phys. Rev.} {\bf D41} (1990) 3405.
%
\bibitem{Dai} Y. B. Dai, C. S. Huang and H. W. Huang,
{\it Phys. Lett.} {\bf B390} (1997) 257, \\ C. S. Huang, L. Wei,
Q. S. Yan and S. H. Zhu, {\it Phys. Rev.} {\bf D63} (2001) 114021.
%
\bibitem{Logan} H. E. Logan and U. Nierste, {\it Nucl. Phys.} {\bf B586}
(2000) 39.
%
\bibitem{Ergur} E. Iltan and G. Turan,
{\it Phys. Rev.} {\bf D63} (2001) 115007.
%
\bibitem{Aliev1} T. M. Aliev and M.Savc{\i}, {\it Phys. Lett.} {\bf B 452} (1999) 318.
%
\bibitem{Choudhury2} S. Rai Choudhury and N. Gaur, {\it hep-ph/0207353}.
%
\bibitem{Babu} K. S. Babu, K. R. S. Balaji and I. Schienbein,{\it Phys. Rev.}, {\bf
D68}, (2003) 014021.
%
\bibitem{Aliev2} T. M. Aliev, V. Bashiry, and M. Savc{\i}, {\it hep-ph/0308069}.
%
\bibitem{Belle} J. Kaneko {\it et al.}, BELLE Collaboration, {\it Phys. Rev. Lett.}, {\bf
90}, (2003) 021801.
%
\bibitem{Ali} A. Ali, E. Lunghi, C. Greub and G. Hiller, {\it Phys. Rev.}, {\bf
D66}, (2002) 034002.
%
\bibitem{HQET}For a review, see, M. Neubert, {\it Phys. Rep.}, {\bf 245}, (1994) 396.
\bibitem{Buchalla} G. Buchalla, A. Buras, and M. Lautenbacher, {\it
Rev. Mod. Phys.}, {\bf 68} (1996) 1125.
%
\bibitem{Wise} B. Grinstein, R. Springer, and M. Wise, {\it
Nucl. Phys.}, {\bf B339} (1990) 269.
%
\bibitem{Buras2} A. J. Buras, M. Misiak, M. M\"{u}nz, and S. Pokorski,
{\it Nucl. Phys.} {\bf B 424} (1994) 372.
%
\bibitem{Misiak} M. Misiak,
{\it Nucl. Phys.} {\bf B 393} (1993) 23; {\bf B 439} (1993) 461 (E).
%
\bibitem{Borzumati} F. Borzumati and C. Greub,
{\it Phys. Rev.} {\bf D 58} (1998) 074004.
%
\bibitem{Ciu} M. Ciuchini, G. Degrassi, P. Gambino, and G. F. Giudice,
{\it Nucl. Phys.} {\bf B 527} (1998) 21.
%
%
\bibitem{Ligeti} Z. Ligeti, I. W. Stewart, M. B. Wise {\it Phys.Lett.} {\bf B420} (1998)  359.
%
\bibitem{Buchalla2} G. Buchalla, G. Hiller and G. Isidori,
{\it Phys. Rev.} {\bf D 63} (2000) 014015.
%
\bibitem{Wolf} L. Wolfenstein,
{\it Phys. Rev. Lett.} {\bf 51} (1983) 1945.
%
\bibitem{AliLunghi} A. Ali, and E. Lunghi,
{\it Eur. Phys. J.} {\bf C 26} (2002) 195.

\end{thebibliography}
\end{document}